\documentclass[sigconf]{acmart}
\usepackage{multirow} 
\usepackage{multicol}
\usepackage{amsmath}
\usepackage{adjustbox}
\usepackage[linesnumbered,ruled,vlined]{algorithm2e}

\AtBeginDocument{%
  \providecommand\BibTeX{{%
    \normalfont B\kern-0.5em{\scshape i\kern-0.25em b}\kern-0.8em\TeX}}}

\setcopyright{acmcopyright}
\copyrightyear{2018}
\acmYear{2018}
\acmDOI{XXXXXXX.XXXXXXX}

\acmConference[Conference acronym 'XX]{Make sure to enter the correct
  conference title from your rights confirmation emai}{June 03--05,
  2018}{Woodstock, NY}
\acmPrice{15.00}
\acmISBN{978-1-4503-XXXX-X/18/06}

\begin{document}

\title{Poisoning Decentralized Collaborative Recommender System and Its Countermeasures}

\author{Ruiqi Zheng}
\email{12150084@mail.sustech.edu.cn}
\affiliation{%
  \institution{Southern University of Science and Technology}
  \city{Shenzhen}
  \country{China}
}

\author{Liang Qu}
\email{qul@mail.sustech.edu.cn}
\affiliation{%
  \institution{Southern University of Science and Technology}
  \city{Shenzhen}
  \country{China}
}

\author{Tong Chen}
\email{tong.chen@uq.edu.au}
\affiliation{%
  \institution{The University of Queensland}
  \city{Brisbane}
  \country{Australia}
}

\author{Kai Zheng}
\email{zhengkai@uestc.edu.cn}
\affiliation{%
  \institution{University of Electronic Science and Technology of China}
  \city{Chengdu}
  \country{China}
}

\author{Yuhui Shi}
\authornote{Corresponding Authors}
\email{shiyh@sustech.edu.cn}
\affiliation{%
  \institution{Southern University of Science and Technology}
  \city{Shenzhen}
  \country{China}
}

\author{Hongzhi Yin}
\authornotemark[1]
\email{db.hongzhi@gmail.com}
\affiliation{%
  \institution{The University of Queensland}
  \city{Brisbane}
  \country{Australia}
}

\begin{abstract}
To make room for privacy and efficiency, the deployment of many recommender systems is experiencing a shift from central servers to personal devices, where the federated recommender systems (FedRecs) and decentralized collaborative recommender systems (DecRecs) are arguably the two most representative paradigms. While both leverage knowledge (e.g., gradients) sharing to facilitate learning local models, FedRecs rely on a central server to coordinate the optimization process, yet in DecRecs, the knowledge sharing directly happens between clients. On the flip side, knowledge sharing also opens a backdoor for model poisoning attacks, where adversaries disguise themselves as benign clients and disseminate polluted knowledge to achieve malicious goals like promoting an item's exposure rate. Although research on such poisoning attacks provides valuable insights into finding security loopholes and corresponding countermeasures, existing attacks mostly focus on FedRecs, and are either inapplicable or ineffective for DecRecs. Compared with FedRecs where the tampered information can be universally distributed to all clients once uploaded to the cloud, each adversary in DecRecs can only communicate with neighbor clients of a small size, confining its impact to a limited range. 

To fill the gap, we present a novel attack method named Poisoning with Adaptive Malicious Neighbors (PAMN). With item promotion in top-$K$ recommendation as the attack objective, PAMN effectively boosts target items' ranks with several adversaries that emulate benign clients (i.e., users) and transfers adaptively crafted gradients conditioned on each adversary's neighbors. 
A diversity-driven regularizer is further designed in PAMN to allow the adversaries to reach a broader group of multifaceted benign users. Moreover, with the vulnerabilities of DecRecs uncovered, a dedicated defensive mechanism based on user-level gradient clipping with sparsified updating is proposed. Extensive experiments demonstrate the effectiveness of the poisoning attack and the robustness of our defensive mechanism.

\end{abstract}

\vspace{-5pt}





\maketitle

\section{Introduction}
Recommender systems, recognized for their effectiveness in assisting users with information filtering, traditionally rely on centralized data collection for training recommendation models \cite{covington2016deep,guo2017deepfm,zhang2021double,wang2017location}. However, the centralized approach poses privacy concerns, potentially violating regulations like the General Data Protection Regulation (GDPR\footnote{https://gdpr-info.eu/}). To address the growing demands for privacy and efficiency, a recent trend involves transferring the model training and deployment of recommender systems from the central servers to users' devices, where users' data can be strictly retained on their local devices \cite{yao2021device,yin2024ondevice, qu2024budgeted}.

Given the sparse nature of local data owned by each client/user, a collaborative learning mechanism \cite{yao2021device,zheng2024decentralized, qu2023semi} is usually in place during training to facilitate the exchange of knowledge (e.g., model gradient) among clients. 
In this context, research can be divided into two main categories: (1) Federated recommender systems (FedRecs) \cite{yang2020federated,chen2018federated,ammad2019federated, qu2024towards} that mainly rely on server-client collaborations. 
Specifically, a central server is responsible for collecting locally trained knowledge, aggregating them using parameter aggregation algorithms (e.g., FedAvg \cite{pmlr-v54-mcmahan17a}), and then redistributing them to all clients. (2) Decentralized collaborative recommender systems (DecRecs) \cite{kermarrec2010RW,defiebre2020decentralized,defiebre2022human,chen2018privacy}, which are based on the client-client collaboration. In this paradigm, the locally trained client models are refined with the knowledge directly obtained from neighboring clients. The dependency on the central server is thus minimized in DecRecs, as its only responsibility is to determine who to communicate with for each client based on user preference similarity.

Despite the benefits of collaborative learning for recommendation, its communicative characteristic inevitably creates vulnerability to attacks, where adversaries 
achieve malicious goals by uploading polluted knowledge \cite{rong2022poisoning}.
Recognizing the susceptibility of recommender system paradigms to potential attacks is crucial to uncovering vulnerabilities within different recommendation paradigms, contributing valuable insights for the development of robust security frameworks \cite{yuan2023manipulating,long2024physical, zhao2023survey}. In this context, poisoning attacks stand out as a representative form for recommendation attacks \cite{yuan2023federated,zhang2021pipattack}, often employed in scenarios such as item promotion to manipulate item exposure rates and rankings in the recommendation list. While various poisoning attacks have been studied extensively, their applicability is predominantly limited to FedRecs, and lacks effectiveness when applied DecRecs. 
As DecRecs emerge as a novel paradigm within recommender systems, its unique security challenges remain insufficiently revealed and explored.


Model poisoning attacks against DecRecs present non-trivial challenges distinct from FedRecs. (1) Firstly, poisoning through knowledge sharing in DecRecs requires adversaries to actively engage in client-to-client communications. Unlike FedRecs, where every client passively participates in knowledge exchange coordinated by the server, communications in DecRecs are primarily enabled between pairwise similar clients. This adds an additional requirement for adversaries to adeptly mimic benign users, thus being qualified as their neighbors in the communication network. 
(2) Secondly, the impact of adversaries in DecRecs is confined to a specific range due to the strong locality of communications between clients. This locality limitation implies that the malicious knowledge shared by adversaries can only influence a small set of neighboring clients, leading to lower attack effectiveness compared with FedRecs where tampered information can be globally disseminated by the server.
As such, to make the poisoning attack effective in DecRecs, adversaries have to maximize their influence by stretching their reach to a diverse set of benign users. 


To unveil the genuine threats posed by poisoning attacks on DecRecs, this paper introduces a novel model poisoning attack method: Poisoning with Adaptive Malicious Neighbors (PAMN). 
Notably, the attack task is the target item promotion, motivated by its high revenue potential and the extensive study it has received within FedRecs \cite{zhang2022pipattack, rong2022fedrecattack, rong2022poisoning}. This paper leverages the well-established objective of poisoning attacks in FedRecs, where the primary objective of PAMN is to enhance the exposure rate of target items, specifically to increase their appearance in more users' top-$K$ recommendation lists.
This is achieved through a collective effort of adversaries generating and transferring adaptive gradients, which are tailored according to the different neighbors with whom the adversaries engage in communication. In light of the above unique challenges in DecRecs, an adversary leverages knowledge collected from its neighbors to update its item embedding, thereby emulating the behavior of benign users. Based on the finding \cite{yuan2023manipulating} that the designed high exposure rate of the target items within adversaries will transfer to benign users, the adversary in our method learns an adaptive user embedding with well-designed item sets and incorporates information from its neighbors' item embeddings. Additionally, a diversity-driven regularizer is applied to all 
adversaries to encourage variety in the data distribution, ensuring diverse benign users are influenced by the model poisoning attack. Subsequently, the local optimization of the target item's rank on the adversaries serves as the basis for the polluted gradient, which is then sent to other users to promote the ranking of target items. By mimicking benign users and boosting the exposure of targeted items, the adversaries' influence can be effectively broadened.


As we expose DecRecs' potential vulnerability to poisoning attacks, in this paper, we move beyond the attack model and additionally address the pressing need for an effective defense mechanism. 
While some studies have explored general collaborative learning \cite{zhang2021survey}. 
These approaches cannot be directly applied to DecRecs due to the non-IID nature of data, which is inherited from FedRecs to DecRecs. Data from different clients are not identically and independently distributed. Users' preferences for the same item may significantly vary. Common byzantine defense methods (e.g., Trimmed Mean \cite{yin2018byzantine}, and Krum \cite{blanchard2017machine}) in collaborative learning, which assumes that clients' data share the same distribution are not applicable. 
Moreover, the absence of a cloud server in DecRecs distinguishes it from FedRecs. Unlike FedRecs, which depends on a central server to aggregate and diminish polluted knowledge from adversaries for defense, DecRecs involve direct communication between clients without the intermediary of a cloud server. 
Consequently, the vulnerabilities in DecRecs demand a unique defense mechanism that considers the non-IID data distribution and the absence of a central server.

In response to these challenges, this paper introduces a novel defense method, User-level Clipping with Sparsified Updating (UCSU), designed specifically for individual user-level defense. Within DecRecs, gradients are used for knowledge exchange instead of direct
sharing the sensitive ratings or interacted items, and users utilize received ingredients to update their local model. For UCSU, all received gradients are clipped to eliminate dominated gradients. Subsequently, clipping with a sparsified updating mechanism is utilized as a defense against model poisoning attacks. The key contributions of this paper can be outlined as follows:
\begin{itemize}
    \item This paper proposes a model poisoning attack method, PAMN, specifically designed to manipulate item ranks within DecRecs. To the best of our knowledge, this work is the first attempt to investigate the vulnerability of DecRecs.
    \item This paper proposes a user-level defense method, UCSU, strategically designed to counteract the threats posed by model poisoning attacks on DecRecs. This defense approach is tailored to the unique characteristics and challenges presented by DecRecs.
    \item Extensive experiments are conducted on two real-world recommendation datasets, providing robust evidence of the effectiveness of the attack method (PAMN) and the robustness of the defense method (UCSU).
\end{itemize}

\vspace{-5pt}
\section{Preliminary}
This section first introduces the framework of the decentralized collaborative recommender systems (DecRecs), and then formulates the model poisoning attack and defense tasks under DecRecs. 

\subsection{Decentralized Collaborative Recommender Systems (DecRecs)}
\label{sec:decrec}

Following \cite{kermarrec2010RW,defiebre2020decentralized,defiebre2022human,chen2018privacy}, this paper adopts the widely recognized framework of DecRecs that users strictly retain sensitive data on their own devices for training local models, and a central server is responsible for assigning neighbors to each client so that they can perform decentralized collaborative learning by exchanging knowledge (e.g., model gradient) via client-client communications. 

In the context of DecRecs, each client manages their sensitive local dataset $\mathcal{O}_i=(u_{i}, v_{ij}, r_{ij})_{j=\{1,...,|\mathcal{O}_i|\}}$, where $v_{ij}$ and $r_{ij}$ are the item that user $u_i$ have interacted with and the corresponding rating, respectively. 
Furthermore, each client also maintains a local model that is typically parameterized by the user's own embedding $\textbf{u}_{i} \in \mathbb{R}^{K}$, all item embeddings $\textbf{V}_{i} \in \mathbb{R}^{K \times I}$, and model parameters $\Theta_{i}$, where $K$ is the dimension of user/item embeddings, and $I$ is the number of items. Since the data on each client is usually limited and insufficient to train an accurate model, DecRecs typically require an additional server that is only responsible for assigning neighbors to each client by calculating similarity scores based on non-sensitive user information such as local item embedding tables \cite{defiebre2022human,defiebre2020decentralized}. In this way, each client can exchange their knowledge (e.g., gradients of item embeddings $\frac{\partial \mathcal{L}_{u_i}}{\partial \textbf{V}_{i}}$) with their neighbors $\mathcal{N}(u_i)$ to optimize local model parameters (e.g., item embedding $\textbf{V}_{i}$) in a client-client collaboration manner defined below:
\begin{equation}
    \textbf{V}_{i} = \textbf{V}_{i} - \gamma Agg(\nabla\textbf{V}_k)_{u_{k} \in \mathcal{N}(u_i)}    
\label{equ:CL}
\end{equation}
where $\nabla\textbf{V}_k = \frac{\partial \mathcal{L}_{u_k}}{\partial \textbf{V}_{k}}$, and $\gamma$ is the learning rate. $Agg(\cdot)$ is the aggregation function (e.g., FedAvg \cite{pmlr-v54-mcmahan17a}), and $\mathcal{L}_{u_{i}}$ is the loss function (e.g., Bayesian Personalized Ranking (BPR) loss \cite{rendle2012bpr}) for the client $u_{i}$.

\subsection{Model Poisoning Attack and Defense}
Although DecRecs can effectively mitigate privacy concerns, the collaborative learning mechanism involving knowledge sharing among clients inevitably makes the system vulnerable to attacks, where polluted knowledge is shared by adversaries.

\textbf{Task 1: Target Item Promotion in DecRecs.} 
Assuming there is a group of users $\widetilde{\mathcal{U}}$ acting as adversaries hidden within decentralized recommender systems, and they share the same set of target items, their goal is to attack the models of benign users $\mathcal{U} \setminus \widetilde{\mathcal{U}}$ during the client-client collaboration process, which only involves exchanging gradients, in order to increase the exposure of target items among benign users. Formally, the goal of the model poisoning attack is to maximize the Exposure Ratio (ER@\textit{K}) \cite{zhang2021pipattack,yuan2023manipulating} that quantifies the extent to which relevant items are adequately represented within the top \textit{K} positions of a ranked list, which is defined as below:
\begin{equation}
 ER@K = \frac{1}{|\widetilde{\mathcal{V}}|} \sum_{v_j \in \widetilde{\mathcal{V}}} \frac{\Bigl|\{{u_i \in \mathcal{U}}| v_j \in \hat{\mathcal{V}_i} \wedge v_j \in \mathcal{V}^{-}_{i}\}\Bigl|}{\Bigl|\{ u_i \in \mathcal{U} | v_j \in \mathcal{V}^{-}_{i} \}\Bigl|}
\label{equ:er@k}
\end{equation}
where $\widetilde{\mathcal{V}}$ is the set of target items, and $\hat{\mathcal{V}_i}$ is the set of recommended items produced by the recommender systems to user $u_i$. $\mathcal{V}^{+}_{i}$ and $\mathcal{V}^{-}_{i}$ represent the set of interacted items and non-interacted items for user $u_i$. Apart from maximizing the exposure rate, the target item promotion should not significantly degenerate the DecRecs' recommendation performance measured by the general recommendation metrics (e.g., Hit Ratio HR@\textit{K} for top-\textit{K} recommendation).




\subsubsection{Prior Knowledge of Attack} This paper minimizes the attack prior knowledge, which assumes that adversaries possess only the knowledge of gradients sent by benign neighbors, aligning with the framework of DecRecs. The attack starts after the initial participation of the cloud central server to identify neighbors. The attackers then compromise certain portions of users as adversaries. 

\textbf{Task 2: Defense Against Model Poisoning Attack.} Unlike FedRecs that have a central server to perform the defense mechanism, the defense for DecRecs is performed on the user side. 
The defense needs to smoothly fit into the standard framework of DecRecs without causing disruptions. Formally, the goal of the defense against model poisoning attacks is to minimize the impact of poisoning attacks, by neutralizing the exposure Ratio (ER@$K$) for the target items, and keeping any potential side effects to a minimum.

\subsubsection{Prior Knowledge of Defense} After the cloud server assigns neighbors to the benign users, the users remain unaware of the identity of other users as either benign or malicious. There is no information available regarding the presence or number of adversaries within their designated neighbors. The users are essentially in the dark about the existence and quantity of adversaries among their neighbors. Moreover, users do not know which items are selected as the target items by the attacker.



\vspace{-5pt}
\section{Proposed Method}
\label{sec:method}

\begin{figure*}[t]
\centering
\includegraphics[width=1\textwidth]{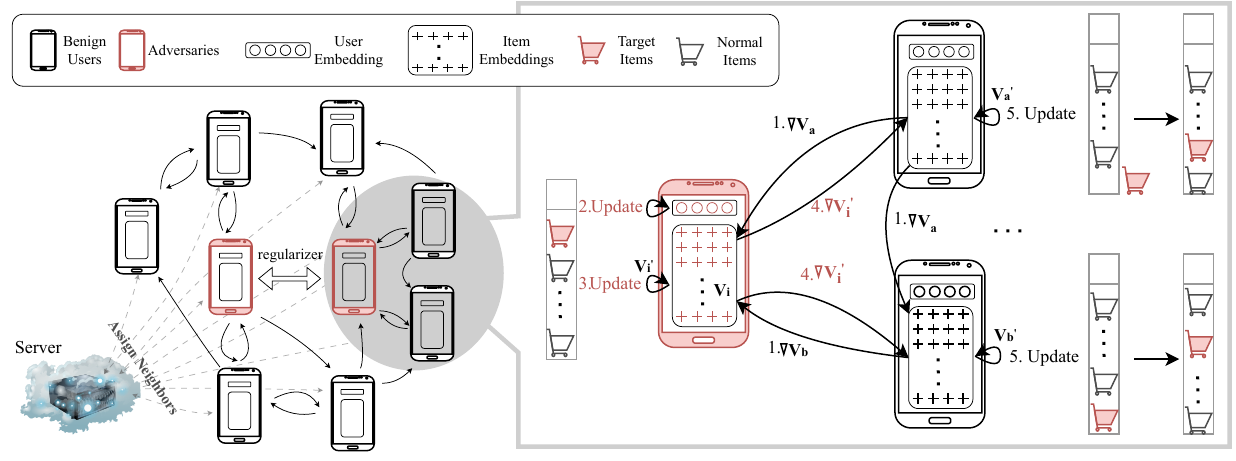} 
\vspace{-2.5em}
\caption{The overview of PAMN. The cloud server assigns users to neighbors who share similar item embedding tables. For the attack, a) The adversary receives gradients from their neighbors to update the user and item embeddings, effectively mimicking the user profiles of these neighbors. b) Adversary fine-tunes item embeddings by optimizing the local ranking of the target items with its fixed user embedding. c) Adversary transfers the gradients of the local item embedding table to promote the ranking of target items for other users within the collaborative network.}
\label{fig:overview}
\vspace{-0.5em}
\end{figure*}

This section provides a comprehensive overview of our poisoning attack method (PAMN) as illustrated in Figure \ref{fig:overview}. The primary elements comprise: (a) The cloud server receives item embeddings from all users, and assigns users' neighbors set $\mathcal{N}(u_i)$ based on the similarity. (b) The adversaries receive gradients from their neighbors and update the user and item embeddings. Then the user embedding is fixed and item embeddings are fine-tuned by optimizing the local ranking of target items. including user mimicking, local item embeddings update, and gradient transferring for item promotion. Furthermore, the defense method (UCSU) is proposed to 
counter the model poisoning attack at the user level for DecRec. Algorithm \ref{alg:total} describes the overall pipeline of integrating PAMN and UCSU with DecRecs. PAMN and UCSU are outlined in Algorithms \ref{alg:attack} and \ref{alg:defense} with corresponding pseudo-code.

\subsection{PAMN: Poisoning with Adaptive Malicious Neighbors}

To streamline the process of identifying neighboring entities and fostering collaborative learning, every client first trains a local model with their local dataset. 
Since the choice of the local model is not a novelty in our attack and defense methods, we use the widely adopted \cite{zhang2021pipattack,yuan2023federated} latent factor recommendation model: NCF \cite{he2017neural}, and describe it in a brief streak.
We utilize a multi-layer feedforward network (FFN) to calculate the estimated rating $\hat{r}_{ij}$:

\begin{equation}
\hat{r}_{ij} = \sigma (\textbf{w} ^\top FFN([\textbf{u}_i, \textbf{v}_j ])),
\label{equ:local}
\end{equation}
where $[\cdot,\cdot]$ is the vector concatenation, $\textbf{w} \in \mathbb{R}^{d_L}$ is the projection weights, and $d_L$ is the dimension of the last feedforward layer. User and item embeddings are optimized by the BPR loss $\mathcal{L}_{u_i}^{loc}(\textbf{u}_i, \textbf{V}_i, \mathcal{V}^{+}_{i}, \mathcal{V}^{-}_{i})$ according to the rating on the interacted and non-interacted item sets.

Then, a server is utilized to assign neighbors to each client by calculating similarity scores based on item embeddings every $\mathcal{T}$ epochs. The high-dimensional item embedding table $\textbf{V}_i$ is reduced to low-dimensional versions while preserving relative distances (e.g., Locality-sensitive hashing \cite{qi2017distributed}) by $\textbf{s}_i = [
Hash(\textbf{v}^{1}_{i}), \cdot\cdot\cdot, Hash(\textbf{v}^{|\mathcal{V}|}_{i})]$, and the similarity is calculated as follows:
\begin{equation}
\begin{aligned}
    Similarity (\textbf{s}_i, \textbf{s}_k) &= \frac{\textbf{s}_i \cdot \textbf{s}_k}{||\textbf{s}_i||||\textbf{s}_k||}
\end{aligned}
\label{equ:similar}
\end{equation}

\begin{algorithm}
\caption{Integration of PAMN and UCSU with DecRec.}
\label{alg:total}
\For{ \textnormal{each} $t = 1,\cdot\cdot\cdot,T$ }{

\If{$t \mod \mathcal{T} = 1$}{
\tcp{server involved in to update neighbors}
    Collect $\{\mathcal{V}_i\}^{|\mathcal{U}|}_{i=1}$ from $u_i \in \mathcal{U}$
    
    Calculate user similarities by Equation \ref{equ:similar}
    
    Send $\mathcal{N}(u_i)$ to every $u_i \in \mathcal{U}$
    }

\For{\textnormal{each} $u_i\in \mathcal{U} $ \textbf{in parallel}}{

\If{$u_i \in \mathcal{U}\setminus \widetilde{\mathcal{U}}$}{

    Execute defense method in Algorithm \ref{alg:defense}
    
    Collect gradients from all neighbors $u_k \in \mathcal{N}(u_i)$
    Update local parameters by Equation \ref{equ:CL} 
    } 
    \Else{Execute attack method in Algorithm \ref{alg:attack} }
}
}
\end{algorithm}

Based on the similarity calculation, the $ N $ users most similar to $ u_i $ are assigned as its neighbors $ \mathcal{N}(u_i)$. Algorithm \ref{alg:total} depicts the neighbor identification process in DecRecs, along with the integration of both attack and defense methods within the DecRecs framework.

Following the prior knowledge of attacks, all adversaries conceal their intention to promote target items before the initial neighbor identification or only activate after that stage. 
To adhere to the client-to-client communication and locality limitation of DecRecs, adversaries need to appear similar to their benign neighbors while promoting target items. The set of $m$ adversaries is denoted as $\widetilde{\mathcal{U}} = \{ \widetilde{u_i}\}^{m}_{i=1} $, and their neighbors set is represented as $\mathcal{N}(\widetilde{u}_i)$. A regularizer (introduced in Section \ref{sec:regu}) is utilized on all adversaries to encourage distinctiveness, expressing diverse user profiles. This diversity allows them the chance to be neighbors to benign users with various preferences as training progresses, covering a broad range of benign users.

The attacks on recommender systems in collaborative learning scenarios can be seen as a special case of backdoor attacks. The goal of PAMN is to maximize the exposure of target items $\widetilde{\mathcal{V}}$. To achieve this, PAMN aims to maximize ER@$K$ in Equation \ref{equ:er@k} by uploading polluted gradients via adversaries in the collaboration. Unfortunately, ER@$K$ is a highly non-differentiable and discontinuous function, leading to difficulties in computing effective poisoned gradients.

\subsubsection{ER@$K$ and Rating Scores.} To address the ER@$K$ discontinuous problem. Predicted scores are employed as a metric to indicate ER@$K$. This is achieved by fostering higher calculated rating values for target items compared to the scores of the top-$K$ recommendation \cite{rong2022poisoning}:

\begin{equation}
\mathcal{L}^{rate} = \sum_{u_i \in \mathcal{U}} \sum_{v_j \in \hat{\mathcal{V}_i} \wedge v_j \notin \widetilde{\mathcal{V}}} \sum_{v_t \in \widetilde{\mathcal{V}} \wedge v_t \notin \mathcal{V}^{+}_{i}} \sigma (\hat{r}_{ij} - \hat{r}_{it})
\label{equ:rating}
\end{equation}
To minimize expression in Equation \ref{equ:rating}, it is necessary to possess information on every normal user representation $\textbf{u}_i$ and clicking histories $\mathcal{V}^{+}_{i}$. However, this is impractical in the context of DecRecs. Additionally, for adversaries to maintain collaboration with benign neighbors, they must appear similar. As a result, this method resorts to approximating the profiles of benign users with those of adversaries.

\subsubsection{Approximation with Adaptive Malicious Neighbors.} 
Referring to the findings in \cite{yuan2023manipulating}, if target items can surface in the top-$K$ recommendations for adversaries, created with randomly selected items, they are more likely to be recommended to benign users by the recommender model. This implies a consistency in the popularity of items between benign users and adversaries. The attack method advances a target item's popularity among benign users through manipulation by our adversaries. During the DecRecs learning stage, when an adversary $\widetilde{u}_{i}$ is involved, it randomly chooses $\alpha$ items (excluding the target items) to form the set of interacted items $\widetilde{\mathcal{V}}^{+}_{i}$ and assembles the local training set $\widetilde{\mathcal{O}_i}$. Equation \ref{equ:rating} is transformed as:

\begin{equation}
\widetilde{\mathcal{L}}^{rate} = \sum_{\widetilde{u_i} \in \widetilde{\mathcal{U}}} \sum_{v_j \in \hat{\widetilde{\mathcal{V}_i}} \wedge v_j \notin \widetilde{\mathcal{V}}} \sum_{v_t \in \widetilde{\mathcal{V}} \wedge v_t \notin \widetilde{\mathcal{V}^{+}_{i}}} \sigma (\hat{r}_{ij} - \hat{r}_{it}),
\label{equ:rating_mal}
\end{equation}
where $\hat{\widetilde{\mathcal{V}_i}}$ is the item set processing highest predicted rating scores for adversary $\widetilde{u_i}$. Attacking goal only depends on $\widetilde{\mathcal{U}}$, $\widetilde{\textbf{V}_i}$, $\widetilde{\Theta_i}$ and $\hat{\widetilde{\mathcal{V}_i}}$. The local $\widetilde{\textbf{V}_i}$ and $\widetilde{\Theta_i}$ can be obtained by the collaborative learning with neighbors (Equation \ref{equ:CL}) for every adversary $\widetilde{u}_i$ in $\widetilde{\mathcal{U}}$:

\begin{equation}
\begin{aligned}
    \widetilde{\Theta_i} &= \widetilde{\Theta_i} - \gamma \frac{1}{|\mathcal{N}(\widetilde{u_i})|}\sum_{u_k \in \mathcal{N}(\widetilde{u_i})} \frac{\partial \mathcal{L}^{loc}_{u_k}}{\partial \Theta_{k}} \\
    \widetilde{\textbf{V}_i} &= \widetilde{\textbf{V}_i} - \gamma \frac{1}{|\mathcal{N}(\widetilde{u_i})|}\sum_{u_k \in \mathcal{N}(\widetilde{u_i})} \frac{\partial \mathcal{L}^{loc}_{u_k}}{\partial \textbf{V}_{k}}
\end{aligned}
\label{equ:CL_Malicious}
\end{equation}

Given that the gradients uploaded for each adversary from its neighbors are different, the adversary strategy is adaptive, tailoring to the specific neighbors. After the adversary $\widetilde{u}_i$ obtains local $\widetilde{\textbf{V}_i}$ and $\widetilde{\Theta_i}$, the private local user embedding for adversaries is awaiting for update. Since the adversaries are all compromised by the attacker, it is rational to assume that attacker has the full control of the adversaries and thus can collect private user embedding from all adversaries.

\subsubsection{Diversified Adversaries Embeddings}
\label{sec:regu}
To encourage distinctiveness, we use a regularizer to express diverse adversaries' profiles. This diversity allows them the chance to be neighbors to benign users with various preferences as training progresses, covering a broad range of benign users. The private adversary's user embedding $\widetilde{\textbf{u}_i}$ is updated by the loss $\widetilde{\mathcal{{L}}}_{user}$, determined by knowledge transferred from the neighbors, locally constructed dataset $\widetilde{\mathcal{O}_i}$ and the diversity regularizer:

\begin{equation}
\begin{aligned}
\widetilde{\mathcal{L}}_{user}  &= \mathcal{L}^{loc}_{\widetilde{u_i} }(\widetilde{\Theta_{i}}, \widetilde{\textbf{V}_i}, \widetilde{\mathcal{O}_i} ) - \lambda \sum_{u_k \in \widetilde{\mathcal{U}}} || \textbf{u}_i - \textbf{u}_k ||_p 
\end{aligned}
\label{equ:local_mal_reg}
\end{equation}
In the regularization term,  $||\cdot||_p$ denotes the $\ell_p$ normalization, which makes distinctive user embeddings to include various information, thus covering a broad range of benign users.

\subsubsection{Substitute Items}
After updating $\widetilde{\textbf{V}_i}$, $\widetilde{\Theta_i}$, and $\widetilde{\textbf{u}}_i$, the recommended item set $\hat{\widetilde{\mathcal{V}}}_i$ can be obtained. Finally, adversary's user embeddings is fixed, and $\widetilde{\Theta_{{i}}}$, $\widetilde{\textbf{V}_i}$ is fine-tuned to minimize $\widetilde{\mathcal{L}}^{rate}$ in Equation \ref{equ:rating_mal}. To further enhance the competitiveness of target items, the attack method aims to improve the prediction scores of the target items by introducing substitute items $\widetilde{\mathcal{V}_i^{si}}$, thereby expanding the pool of competitive items. The selection of substitute items is carried out by adversary $\widetilde{u_i}$ utilizing the item embedding table $\widetilde{\mathcal{V}_{i}}$. In this process, $\widetilde{u_i}$ picks items with both higher item embedding similarity to the target items and relatively higher preference scores, effectively augmenting the competition item set for user $\widetilde{u_i}$. Therefore Equation \ref{equ:rating_mal} is modified as:

\begin{equation}
\widetilde{\mathcal{L}}^{rate}_{si} = \sum_{\widetilde{u_i} \in \widetilde{\mathcal{U}}} \sum_{v_j \in \hat{\widetilde{\mathcal{V}_i}} \wedge v_j \notin \widetilde{\mathcal{V}}} \sum_{v_t \in \{ \widetilde{\mathcal{V}} \cup \widetilde{\mathcal{V}_i^{si}}\} \wedge v_t \notin \widetilde{\mathcal{V}^{+}_{i}}} \sigma (\hat{r}_{ij} - \hat{r}_{it})
\label{equ:rating_mal_ap}
\end{equation}

$\widetilde{\Theta_{{i}}}$, and $\widetilde{\textbf{V}_i}$ are fine-tuned by the gradient on the loss function $\widetilde{\mathcal{L}}_{rate}$ in Equation \ref{equ:rating_mal_ap}. The gradient $\nabla \widetilde{\Theta_{{i}}}$ and $\nabla \widetilde{\textbf{V}_i}$ from the adversary are transferred to neighbors to promote the target items. Algorithm \ref{alg:attack} describes the attack method in pseudo-code.

\vspace{-5pt}
\begin{algorithm}
\caption{Optimizing PAMN.}
\label{alg:attack}
\tcp{on adversaries' side}
\For{\textnormal{each} $\widetilde{u_i}\in \widetilde{\mathcal{U}}$ \textbf{in parallel}}{
Randomly construct training set $\widetilde{\mathcal{O}}_i$

Collect $\nabla \textbf{V}_k$ and $\nabla \Theta_k$ from neighbor $u_k \in \mathcal{N}(\widetilde{u_i})$

Update $\widetilde{\textbf{V}_i}$ and $\widetilde{\Theta_i}$ with Equation \ref{equ:CL_Malicious}

Update adversary's user embedding $\widetilde{\textbf{u}}_i$ by taking a gradient step w.r.t $\widetilde{\mathcal{L}}_{user}$ in Equation \ref{equ:local_mal_reg}

Update  $\widetilde{\textbf{V}_i}$ and $\widetilde{\Theta_{{i}}}$ by taking a gradient step w.r.t $\widetilde{\mathcal{L}}^{rate}_{si}$ in Equation \ref{equ:rating_mal_ap}

Transfer  $\nabla \widetilde{\textbf{V}_i}$ and $\nabla \widetilde{\Theta_{{i}}}$ to all neighbors in $\mathcal{N}(\widetilde{u_i})$}
\end{algorithm}

\subsubsection{Discussion}

This section systematically compares the technical rationale of our proposed method for DecRecs with existing attack methods for FedRecs.

Both A-hum \cite{rong2022poisoning} and PSMU \cite{yuan2023manipulating} employ a mechanism to create synthetic users that emulate benign users. They utilize a substitute item set to amplify the impact of target items. However, in FedRecs scenarios, the cloud server randomly selects users for aggregation and then distributes the aggregated information to all users, treating all users uniformly. This approach disregards the inherent diversity in user preferences in recommendation systems. A-hum and PSMU employ a uniform strategy for constructing adversaries, regardless of the number of adversaries.

Implementing A-hum and PSMU in DecRecs scenarios is impractical since, in the initial stage of neighbor identification, adversaries are assigned to users with similar preferences, constituting a small portion of users. As the target items are unpopular in the systems, adversaries lack the opportunity to significantly influence other benign users with diverse preferences. In contrast, PAMN adopts an adaptive approach to constructing adversaries. When two adversaries have different neighbors, the constructed ones are distinct. Notably, our approach introduces a diversity-driven regularizer, which further alleviates the limitations imposed by locality.

\subsection{UCSU: User-level Clipping with Sparsified Updating for Defense}
The susceptibility of DecRecs to poisoning attacks underscores the critical necessity for an innovative defense mechanism. Unfortunately, current research has overlooked this security concern within DecRecs, which, distinct from FedRecs, lacks a centralized server to aggregate and mitigate polluted knowledge from adversaries.

In response, this paper proposes a paradigm shift in defense strategy, relocating the defense mechanism from the cloud server to the user level. This approach termed User-level Clipping with Sparsified Updating for Defense (UCSU), is specifically designed to address the unique challenges posed by DecRecs. Notably, UCSU exclusively processes gradients related to item embeddings. This focused approach is justified by the fact that a substantial amount of benign users effectively impede the spreading of the polluted gradients of model parameters $\Theta$.

\begin{algorithm}
\caption{Optimizing UCSU.}
\label{alg:defense}

\tcp{on benign users' side}
\For{\textnormal{each} $u_i\in \mathcal{U} \setminus \widetilde{\mathcal{U}}$ \textbf{in parallel}}{

Collect $\nabla \textbf{V}_k$ from all neighbors $u_k \in \mathcal{N}(u_i)$

Update $\nabla \textbf{V}_k$ with gradient clip in Equation \ref{equ:gradient_clipping}

Update memory bank $\mathcal{B}_i = \mathcal{B}_i \cup \{\nabla \textbf{V}_k\}$ 

Pop-out top $\nu$ percent of gradients $\{\nabla \textbf{V}_k \}^{\nu}_{k=1}$

Output adaptive gradient $\nabla \textbf{V}^{'}_k$ with Equation \ref{equ:gradient_clipping_avg}
}

\end{algorithm}

\subsubsection{ User-level Clipping} Inspired by the gradient clipping in FedRecs \cite{yuan2023manipulating}, our approach involves manipulating benign users to clip all received gradients from their neighbors. This strategic intervention ensures that the adaptive malicious neighbors of benign users can only transfer limited polluted knowledge, thereby restricting their ability to promote target items: 
\begin{equation}
\frac{1}{|\mathcal{N}({u_i})|}\sum_{u_k \in \mathcal{N}(u_i)} \nabla \textbf{V}_k = \frac{1}{|\mathcal{N}({u_i})|}\sum_{u_k \in \mathcal{N}(u_i)} \nabla \textbf{V}_k \cdot \min \left( 1, \frac{\mu}{||\nabla \textbf{V}_k||_p}\right)
\label{equ:gradient_clipping}
\end{equation}
where $\nabla \textbf{V}_k$ is the gradients transferred by the neighbors of the benign user $u_i$. The magnitude of poisoned gradients is restricted by number of adversaries present within the designated neighbor set: 
\begin{equation}
\left\lVert \sum_{\widetilde{u_k} \in \{\mathcal{N}(u_i) \cap \widetilde{\mathcal{U}} \} } \nabla \widetilde{V}_k \right\rVert _p \leq \mu \left| \mathcal{N}(u_i) \cap \widetilde{\mathcal{U}} \right|
\label{}
\end{equation}
However, the attack can still be effective by directly increasing the number of malicious neighbors.

\subsubsection{User-level Clipping with Sparsified Updating}
To further reduce the influence of attack method, all the received gradients are stored in the memory bank $\mathcal{B}_i$ for benign user $u_i$. For each collaborative learning, only unitize $\nu$ percent of gradients with the largest value from the bank, and clean all other gradients. Furthermore, the $\mu$ in Equation \ref{equ:gradient_clipping}  is modified as the average normalization of the gradients for a more flexible and adaptive updating:

\begin{equation}
 \nabla \textbf{V}^{'}_k = \nabla \textbf{V}_k \cdot \min \left( 1, \frac{\text{avg} ||\nabla \textbf{V}_k||_p}{||\nabla \textbf{V}_k||_p}\right)
\label{equ:gradient_clipping_avg}
\end{equation}

In most cases, the number of adversaries is lower than that of benign users. Consequently, the magnitude of poisoned gradients tends to be relatively small in the initial stages of the attack, reducing the likelihood of their selection as top gradients and thereby delaying the overall attack. The effectiveness of these poisoned gradients relies on waiting until the accumulative gradient magnitudes of the target items reach a significant level. However, throughout this accumulation process, the gradients from benign users gain progressively higher probabilities of diluting the impact of these poisoned gradients.

\section{Experiments}

To validate the effectiveness of the attack method PAMN and defense method UCSU, experiments are conducted  to answer the following research questions (RQs):

\begin{itemize}
    \item \textbf{RQ1}: How does our proposed attack method perform compared with other poisoning attack baselines?
    \item \textbf{RQ2:} How does our proposed defense method perform against poisoning attacks compared with other defense baselines?
    \item \textbf{RQ3:} How do different  parts of our method contribute the final performance?
    \item \textbf{RQ4:} How does the number of adversaries influence the overall performance?

\end{itemize}

\subsection{Datasets}
Following \cite{yuan2023federated,zhang2021pipattack}, the experiments employ two widely used recommendation datasets, namely MovieLens-1M \cite{harper2015movielens} and Amazon Digital Music \cite{mcauley2015image} as shown in Table \ref{tab:dataset}. The datasets and the pre-processing techniques are introduced as follows:

\begin{itemize}
\item \textbf{MovieLens-1M}: This dataset comprises user-application interactions, specifically user ratings assigned to movies. It encompasses a total of 1,000,208 ratings, 3,706 movies and 6,040 users.

\item \textbf{Amazon Digital Music}: This dataset captures user ratings for music items available on Amazon. It encompasses 169,781 interactions between 11,797 products and 16,566 users.

\end{itemize}

Following \cite{rong2022poisoning,zhang2021pipattack,yuan2023manipulating}, a filtering process is implemented to exclude items and users with less than five interactions. Subsequently, the user-item ratings are binarized, transforming all ratings to $r_{ij} = 1$, with negative instances sampled at a ratio of 1:4. The dataset is then partitioned, with 80\% of the data allocated to the training set and the remaining 20\% designated for the test set.

\begin{table}[htbp]
 
  \centering
  \caption{The statistics of datasets.}
  \vspace{-8pt}
    \begin{tabular}{c|c|c}
    \toprule
    Dataset & MovieLens-1M & Digital Music  \\
    \midrule
    \midrule
    \#User & 6,040 & 16,566  \\
    \#Item & 3,900 & 11.797  \\
    \#Interactions & 1,000,208 & 169,781  \\
    \bottomrule
    \end{tabular}%
  \label{tab:dataset}%

  \vspace{-10pt}
\end{table}%

\subsection{Evaluation Protocols}

Our approach aims to enhance the authenticity of gradient poisoning threats by minimizing prior knowledge. In this context, adversaries are assumed to possess solely the knowledge of gradients transmitted by benign neighbors, a premise consistent with the DecRecs framework. The attack commences after the initial involvement of the cloud central server in identifying neighbors and user groups. Subsequently, a specific number of users are compromised, and marked as adversaries.

Regarding recommendation metrics, two popular metrics are employed to assess the effectiveness of the attack and defense methods \cite{yuan2023manipulating,zhang2021pipattack}: (1) Exposure Rate (ER@20)  quantifies the extent to which target items are adequately represented within the top 20 positions of a ranked list. (2) Hit Rate (HR@20) represents the fraction of the top 20 recommended items that are in a set of items relevant to the user.

An ideal targeted attack method should enhance the exposure rate while minimizing adverse effects on the overall performance of DecRecs. As for the defense method, it should prevent the attack effect while minimizing any negative impact on DecRecs.

\subsection{Baselines}
As the first work investigating modeling poisoning attacks and their countermeasures for DecRecs, there is a challenge of lacking existing benchmarks or baselines for comparison. In the absence of prior work applicable as baselines, general poisoning attack and defense methods designed for FedRecs and collaborative learning in recommendation systems are explored and then modified to suit the DecRecs scenarios for a fair and meaningful comparison.
\subsubsection{Attack Baselines} It's crucial to note that certain attacks \cite{gunes2014shilling,zhang2021pipattack} require knowledge of sensitive information about benign users, such as the items they have clicked before. This requirement disrupts the DecRecs pipeline and renders them infeasible for implementation in attack scenarios. After a thorough review, the following baselines are selected:
\begin{itemize}
\item \textbf{No Attack:} It represents the normal ER@$K$ of target items and showcases the performance of DecRecs without attacking.

\item \textbf{Random Attack (RA)} \cite{kapoor2017review}: A straightforward data poisoning attack technique with randomly selected interactions including target items.

\item \textbf{Explicit Boosting (EB)} \cite{zhang2021pipattack}: A model poisoning attack for FedRecs that operates independently of sensitive knowledge from benign users.

\item \textbf{PSMU} \cite{yuan2023manipulating}: The SOTA gradient poisoning attack for FedRecs, which is adapted to suit the DecRecs paradigm.
\end{itemize}

\subsubsection{Defense Baselines} Regrettably, existing research has not addressed this security issue in DecRecs. While some studies have explored general collaborative learning \cite{zhang2021survey} and FedRecs \cite{sun2022survey}, these approaches cannot be directly applied to DecRecs. Modifications are made on the following defense method for a fair comparison. 
\begin{itemize}
    \item \textbf{No Defense:} This method evaluates the original DecRecs' performance under specific attacks.
    
    \item \textbf{Median} \cite{yin2018byzantine}: It aggregates all received gradient values and output the mean of them as the result.
    
    \item \textbf{Trimmed Mean} \cite{yin2018byzantine}: This defense strategy aggregates gradients by excluding the smallest and largest values of a parameter and taking the mean operation on the left gradients.
    
    \item \textbf{Item-based Krum} \cite{blanchard2017machine}: The original Krum is inapplicable to DecRecs due to the incomparability of uploaded gradients from different clients. To adapt Krum to our problem setting, Item-level Krum is modified according to  \cite{yuan2023manipulating}. The aggregated gradient for each item is chosen by selecting the one that is closest to the average of the received gradients.
    
    \item \textbf{$\ell_2$ Gradient Clipping:} Inspired by the defense method in FedRecs \cite{yuan2023manipulating}, this defense approach has been modified from the cloud-server to the user's side to align with DecRecs.
\end{itemize}

\subsection{Implementation Details}

Both the attack method (PAMN) and defense method (UCSU) are designed for general DecRecs. Without loss of generality, we implement NCF \cite{he2017neural} with a three-layer feedforward network (dimensions: 64, 32, 16), as the base model for the local user recommender systems. The DecRecs framework is implemented following \cite{chen2018privacy}. The size of the user and item embedding is determined as 32. Adam optimizer is adopted with a learning rate $\gamma = 0.001$. For adaptive malicious neighbors, we set $\alpha = 30$ to construct the local training set, and the neighbor number $N = 50$. In the defense mechanism, we select $\nu = 10\%$ for the gradient bank, and $\mu = 1$. $p=2$ represents $\ell_2$ normalization throughout the paper. The proportion of adversaries is $\xi = 1\%$ if no specification. We discuss the influence of the number of adversaries in Section \ref{sec:number}.

\begin{figure*}
\begin{minipage}[t]{0.67 \columnwidth}
  \includegraphics[width=\linewidth]{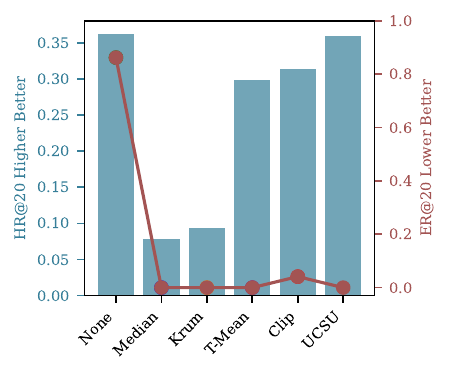}
    \vspace{-18pt}
  \caption{The performance defense method UCSU and defense baselines on MovieLens dataset, represented by HR@20 and ER@20.}
  \label{fig:defense}
\end{minipage}\hfill 
\begin{minipage}[t]{0.67\columnwidth}
  \includegraphics[width=\linewidth]{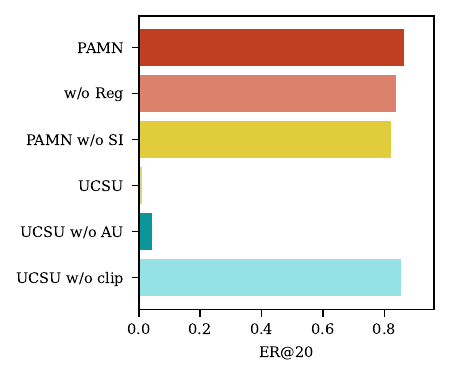}
    \vspace{-18pt}
  \caption{Result of ablation experiment on different
variants of PAMN and UCSU on MovieLens dataset.}
  \label{fig:ablation}
\end{minipage}\hfill
\begin{minipage}[t]{0.67\columnwidth}
  \includegraphics[width=\linewidth]{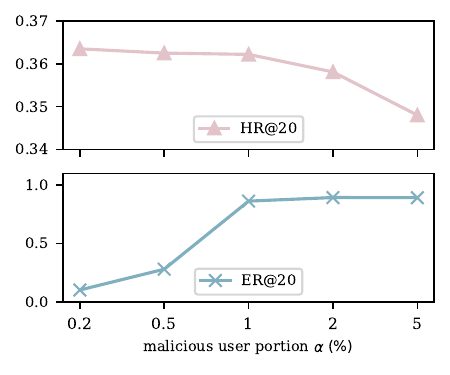}
    \vspace{-18pt}
  \caption{PAMN  performance for different numbers
of malicious neighbors on MovieLens dataset, represented by HR@20 and ER@20.}
  \label{fig:hyper}
\end{minipage}
\vspace{-2pt}
\end{figure*}

\subsection{Attack Performance towards DecRecs (RQ1)}

This section assesses the effectiveness of the proposed attack method in the context of the top-$K$ recommendation task, comparing it with other attack baselines. Experimental results are presented in Table \ref{tab:RQ1},highlighting key observations:

For both the MovieLens and Amazon Music datasets, the implementation of all four attack methods shows no significant impact on the performance of the recommender systems. This adherence to stealthiness is a fundamental requirement for any attack approach. 

Classical attack methods for federated recommender systems, namely RA and EB, fail to effectively promote unpopular items in the recommender systems. This direct implementation of these attack methods in the DecRecs scenario proves unfeasible.

PSMU exhibits superior attack performance compared to RA and EB. One potential explanation is that PSMU not only promotes the target items but also includes the promotion of similar items additionally. This contributes to the establishment of a preference trend towards target items. However, the uniform attack strategy for all adversaries in PSMU results in limited communication with benign users, affecting overall attack performance.

The proposed method, PAMN, achieves the best attack performance compared to all baselines. This is attributed to its adaptive response to diverse neighbors with whom adversaries communicate. Furthermore, the encouragement of user diversity enhances the likelihood of communication with a broad range of benign users.

\begin{table}[htbp]
  \centering
  \caption{The performance of attack methods.}
\vspace{-10pt}
\begin{tabular}{c|cc|cc}
\toprule
\multirow{2}[4]{*}{Methods} & \multicolumn{2}{c|}{MovieLens} & \multicolumn{2}{c}{Amazon Music} \\
\cmidrule{2-5}      & \multicolumn{1}{c|}{HR@20} & ER@20 & \multicolumn{1}{c|}{HR@20} & ER@20 \\
\midrule
\midrule
No Attack & 0.3635 & 0.0000 & 0.0126 & 0.0000 \\
RA    & 0.3531 & 0.0000 & 0.0113 & 0.0000 \\
EB    & 0.3581 & 0.0000 & 0.0120 & 0.0000 \\
PSMU  & 0.3495 & 0.1478 & 0.0118 & 0.0271 \\
Proposed & 0.3622 & 0.8627 & 0.0124 & 0.9168 \\
\bottomrule
\end{tabular}%

\label{tab:RQ1}
\vspace{-10pt}
\end{table}%
\vspace{-10pt}

\subsection{Defense Performance towards Attack (RQ2)}


The efficacy of the PAMN attack method underscores the vulnerability of DecRecs. To assess the feasibility of our proposed defense method UCSU, it is implemented on all benign users when sharing knowledge between users. According to prior knowledge of defense, after the cloud server assigns
neighbors to the benign users, they remain unaware of the identity
of other users as either benign or malicious. There is no information
available regarding the presence or number of adversaries
within their designated neighbors. The users are essentially in the
dark about the existence and quantity of malicious entities among
their neighbors. Therefore, the defense method will be activated from the beginning. The results are presented in Figure \ref{fig:defense}, illustrating both the recommendation and attack performances.

With defense as the primary objective, we aim to achieve two key outcomes: first, minimize the exposure rate of target items to diminish the effectiveness of the attack, and second, avoid compromising recommendation performance to maintain an effective recommendation model. Consequently, a higher HR@20 and a lower ER@20 are desirable. 

 While Median and Krum effectively reduce the impact of the attack, they come at the cost of a significant decrease in recommendation performance, which is deemed unacceptable in recommendation scenarios. The Clip method maintains a relatively modest decrease in recommendation performance, but its ability to counteract the attack is not fully effective. UCSU achieves a negligible decrease in recommendation performance while effectively mitigating the impact of the attack. This dual achievement positions UCSU as a robust and effective defense mechanism.

\subsection{Ablation Study (RQ3)}
The section focuses on the impact of key components in our proposed diversity-driven regularization, substitute items for the attack method, and gradient clipping with adaptive updating for the defense method. Specifically, we implement the proposed methods without one component while keeping the other components unchanged, conducting experiments on the MovieLens dataset, with similar trends observed in other datasets. The results are presented in Figure \ref{fig:ablation}. Key observations include:

\textit{PAMN w/o Reg} removes the regularization term for adversaries. The attack performance decreases, suggesting that without regularization, adversaries tend to exhibit similar user embeddings. This limits their opportunity to communicate with benign users with diverse preferences and promote target items for them.

    \textit{PAMN w/o SI} removes substitute items, which are included in the target item sets to amplify competition. The decreased effectiveness emphasizes the importance of substitute items in the attack method.

    \textit{UCSU w/o AU} deletes adaptive updating for the defense method is removed. The exposure rate of target items increases, indicating a degradation of the defense method.

    \textit{UCSU w/o clip} deletes user-level gradient clipping from the defense method, rendering the defense method ineffective. One possible explanation is that without clipping, gradients for unpopular target items are dominated by adversaries.

\subsection{Adversary Amount Study (RQ4)}
\label{sec:number}

To explore the influence of varying numbers of adversaries on the proposed method PAMN, we vary the proportion of adversaries $\xi$ (in percentage), ranging between $\{0.2, 0.5, 1, 2, 5\}$. The resulting recommendation outcomes, measured by HR@20, and the attack performance, assessed by ER@20, are documented in Figure \ref{fig:hyper}. 

Starting with an adversary proportion of 0.2\%, the recommendation performance is comparable to that achieved without any attack. However, the exposure rate of target items remains low. An explanation is that adversaries have a certain number of neighbors, and limited adversaries cannot widely spread the trend of target items among all benign users. As training progresses, all users learn the actual item embeddings, rendering the attack less effective.

As the proportion increases to 1\%, the recommendation performance stabilizes while the promotion of target items rises. This indicates a balance between spreading the polluted gradients and benign gradients within the decentralized learning framework. When the proportion reaches 5\%, the recommendation performance notably decreases. This could be attributed to an excess of polluted gradients with no actual knowledge hindering the effectiveness of the decentralized collaborative recommender systems.

\section{Related Work}

\subsection{On-device Recommender Systems}

On-device Recommender Systems include two primary approaches. One is on-device deployment \cite{2020Next, xia2023efficient}, where the model is entirely trained on the cloud side and deployed the recommender systems on the device side for timely generation of recommendation results. Another approach is on-device learning \cite{2020Distributed,2021PREFER}, where the recommendation model is trained fully on the client side without uploading sensitive user embedding to the cloud server. Collaborative learning (CL) \cite{2021PREFER, 2022YE} has been introduced to supplement the scarcity of local user-item interactions. This promotes knowledge sharing among users, which utilizes the knowledge from individual users, and does not expose the local sensitive data. CL-based recommender systems can be further classified into: federated recommender systems (FedRecs) \cite{kermarrec2010RW,defiebre2020decentralized,defiebre2022human}, and decentralized collaborative recommender systems (DecRecs) \cite{chen2018privacy,2022Decentralized}.

\subsection{Decentralized Collaborative Recommender Systems (DecRecs)}
Unlike FedRecs, which require a central server all the time to collect, aggregate, and then distribute the public models to users, DecRecs \cite{chen2018privacy,2022Decentralized,yang2022dpmf,long2023model} allocate a distinct role to the central server in the initial phase. The central server only assigns users to neighbors who share similar user preferences. Then, user local models undergo refinement by local optimization and communication with  neighbors.  
DMF \cite{chen2018privacy} is the first attempt to implement Matrix Factorization (MF) in decentralized learning for sequential recommendation. Neighbors are identified based on the geographical information. DPMF \cite{yang2022dpmf} employs probabilistic MF to capture item characteristics and user preferences using both implicit and explicit feedback.
DCLR \cite{2022Decentralized} takes a step further by introducing semantic and geographic neighbors for Point-of-Next-Interest (POI) recommendation. Users collaboratively learn with nearby users due to the high likelihood of visiting the same next location. MAC \cite{long2023model} implements a mechanism for neighborhood identification at the user level during training. Beyond the assignment and re-assignment of neighbors by the cloud server, users have the autonomy to choose to communicate with high-quality neighbors contributing meaningful knowledge.

\subsection{Attacks for Federated Recommender Systems (FedRecs)}

In recent times, collaborative learning frameworks for recommendations have faced a surge in security and privacy threats. These encompass various attacks, such as data poisoning attacks \cite{tolpegin2020data, wu2022fedattack}, model poisoning attacks \cite{zhang2021pipattack, rong2022fedrecattack, rong2022poisoning}, and inference attacks \cite{nasr2019comprehensive, yuan2023interaction}.

Data poisoning injects fake item interactions or intentionally mislabeled data \cite{tolpegin2020data} into the training dataset. FedAttack \cite{wu2022fedattack} disrupts FedRecs with limited Byzantine clients. 
For model poisoning attacks, PipAttack \cite{zhang2022pipattack} is the first attempt on FedRecs, with the objective of enhancing the exposure chances of target items. FedRecAttack \cite{rong2022fedrecattack} accomplishes similar item promotion goals through adversary interventions. A-hum \cite{rong2022poisoning} focuses specifically on model poisoning attacks on FedRecs without prior knowledge of the users' interacted items.
Shifting to inference attacks, Nasr et al. \cite{nasr2019comprehensive} conducted a comprehensive study on class-level membership inference attacks in general federated learning framework under both white-box and black-box settings. Yuan et al. \cite{yuan2023interaction} devised an interaction-level inference attack to reveal the sensitive personal clicking histories by their uploaded models.

\section{Conclusion}

This paper explored the vulnerabilities and security challenges associated with decentralized recommender systems (DecRecs). The shift from centralized to on-device recommendation systems, driven by privacy concerns and efficiency demands, introduced new risks, particularly in the form of model poisoning attacks. The paper introduced a novel model poisoning attack method, Poisoning with Adaptive Malicious Neighbors (PAMN), specifically designed for DecRecs, and proposed a user-level defense mechanism, User-level Clipping with Sparsified Updating (UCSU), tailored to address the unique challenges of DecRecs. Experiments conducted on two real-world datasets demonstrated the generalization and effectiveness of both the attack and defense methods, contributing valuable insights to enhance the security of DecRecs.

\section*{Acknowledgement}
This work is partially supported by the National Key R\&D Program of China under the Grant No. 2023YFE0106300 and 2017YFC0804002, Australian Research Council under the streams of Future Fellowship (Grant No. FT210100624), Discovery Early Career Researcher Award (Grant No. DE230101033), Discovery Project (Grants No. DP240101108, and No. DP240101814), Shenzhen Fundamental Research Program under the Grant No. JCYJ20200109141235597, and National Science Foundation of China under Grant No. 62250710682 and 61761136008.

\bibliographystyle{ACM-Reference-Format}
\bibliography{main}


\begin{thebibliography}{49}


\ifx \showCODEN    \undefined \def \showCODEN     #1{\unskip}     \fi
\ifx \showDOI      \undefined \def \showDOI       #1{#1}\fi
\ifx \showISBNx    \undefined \def \showISBNx     #1{\unskip}     \fi
\ifx \showISBNxiii \undefined \def \showISBNxiii  #1{\unskip}     \fi
\ifx \showISSN     \undefined \def \showISSN      #1{\unskip}     \fi
\ifx \showLCCN     \undefined \def \showLCCN      #1{\unskip}     \fi
\ifx \shownote     \undefined \def \shownote      #1{#1}          \fi
\ifx \showarticletitle \undefined \def \showarticletitle #1{#1}   \fi
\ifx \showURL      \undefined \def \showURL       {\relax}        \fi
\providecommand\bibfield[2]{#2}
\providecommand\bibinfo[2]{#2}
\providecommand\natexlab[1]{#1}
\providecommand\showeprint[2][]{arXiv:#2}

\bibitem[Ammad-Ud-Din et~al\mbox{.}(2019)]%
        {ammad2019federated}
\bibfield{author}{\bibinfo{person}{Muhammad Ammad-Ud-Din}, \bibinfo{person}{Elena Ivannikova}, \bibinfo{person}{Suleiman~A Khan}, \bibinfo{person}{Were Oyomno}, \bibinfo{person}{Qiang Fu}, \bibinfo{person}{Kuan~Eeik Tan}, {and} \bibinfo{person}{Adrian Flanagan}.} \bibinfo{year}{2019}\natexlab{}.
\newblock \showarticletitle{Federated collaborative filtering for privacy-preserving personalized recommendation system}.
\newblock \bibinfo{journal}{\emph{arXiv preprint arXiv:1901.09888}} (\bibinfo{year}{2019}).
\newblock


\bibitem[Bistritz et~al\mbox{.}(2020)]%
        {2020Distributed}
\bibfield{author}{\bibinfo{person}{I. Bistritz}, \bibinfo{person}{A. Mann}, {and} \bibinfo{person}{N. Bambos}.} \bibinfo{year}{2020}\natexlab{}.
\newblock \showarticletitle{Distributed Distillation for On-Device Learning}. In \bibinfo{booktitle}{\emph{Neural Information Processing Systems}}.
\newblock


\bibitem[Blanchard et~al\mbox{.}(2017)]%
        {blanchard2017machine}
\bibfield{author}{\bibinfo{person}{Peva Blanchard}, \bibinfo{person}{El~Mahdi El~Mhamdi}, \bibinfo{person}{Rachid Guerraoui}, {and} \bibinfo{person}{Julien Stainer}.} \bibinfo{year}{2017}\natexlab{}.
\newblock \showarticletitle{Machine learning with adversaries: Byzantine tolerant gradient descent}.
\newblock \bibinfo{journal}{\emph{Advances in neural information processing systems}}  \bibinfo{volume}{30} (\bibinfo{year}{2017}).
\newblock


\bibitem[Chen et~al\mbox{.}(2018a)]%
        {chen2018privacy}
\bibfield{author}{\bibinfo{person}{Chaochao Chen}, \bibinfo{person}{Ziqi Liu}, \bibinfo{person}{Peilin Zhao}, \bibinfo{person}{Jun Zhou}, {and} \bibinfo{person}{Xiaolong Li}.} \bibinfo{year}{2018}\natexlab{a}.
\newblock \showarticletitle{Privacy preserving point-of-interest recommendation using decentralized matrix factorization}. In \bibinfo{booktitle}{\emph{Proceedings of the AAAI Conference on Artificial Intelligence}}, Vol.~\bibinfo{volume}{32}.
\newblock


\bibitem[Chen et~al\mbox{.}(2018b)]%
        {chen2018federated}
\bibfield{author}{\bibinfo{person}{Fei Chen}, \bibinfo{person}{Mi Luo}, \bibinfo{person}{Zhenhua Dong}, \bibinfo{person}{Zhenguo Li}, {and} \bibinfo{person}{Xiuqiang He}.} \bibinfo{year}{2018}\natexlab{b}.
\newblock \showarticletitle{Federated meta-learning with fast convergence and efficient communication}.
\newblock \bibinfo{journal}{\emph{arXiv preprint arXiv:1802.07876}} (\bibinfo{year}{2018}).
\newblock


\bibitem[Covington et~al\mbox{.}(2016)]%
        {covington2016deep}
\bibfield{author}{\bibinfo{person}{Paul Covington}, \bibinfo{person}{Jay Adams}, {and} \bibinfo{person}{Emre Sargin}.} \bibinfo{year}{2016}\natexlab{}.
\newblock \showarticletitle{Deep neural networks for youtube recommendations}. In \bibinfo{booktitle}{\emph{Proceedings of the 10th ACM conference on recommender systems}}. \bibinfo{pages}{191--198}.
\newblock


\bibitem[Defiebre et~al\mbox{.}(2020)]%
        {defiebre2020decentralized}
\bibfield{author}{\bibinfo{person}{Daniel Defiebre}, \bibinfo{person}{Dimitris Sacharidis}, {and} \bibinfo{person}{Panagiotis Germanakos}.} \bibinfo{year}{2020}\natexlab{}.
\newblock \showarticletitle{A decentralized recommendation engine in the social internet of things}. In \bibinfo{booktitle}{\emph{Adjunct Publication of the 28th ACM Conference on User Modeling, Adaptation and Personalization}}. \bibinfo{pages}{77--82}.
\newblock


\bibitem[Defiebre et~al\mbox{.}(2022)]%
        {defiebre2022human}
\bibfield{author}{\bibinfo{person}{Daniel Defiebre}, \bibinfo{person}{Dimitris Sacharidis}, {and} \bibinfo{person}{Panagiotis Germanakos}.} \bibinfo{year}{2022}\natexlab{}.
\newblock \showarticletitle{A human-centered decentralized architecture and recommendation engine in SIoT}.
\newblock \bibinfo{journal}{\emph{User Modeling and User-Adapted Interaction}} \bibinfo{volume}{32}, \bibinfo{number}{3} (\bibinfo{year}{2022}), \bibinfo{pages}{297--353}.
\newblock


\bibitem[Gunes et~al\mbox{.}(2014)]%
        {gunes2014shilling}
\bibfield{author}{\bibinfo{person}{Ihsan Gunes}, \bibinfo{person}{Cihan Kaleli}, \bibinfo{person}{Alper Bilge}, {and} \bibinfo{person}{Huseyin Polat}.} \bibinfo{year}{2014}\natexlab{}.
\newblock \showarticletitle{Shilling attacks against recommender systems: a comprehensive survey}.
\newblock \bibinfo{journal}{\emph{Artificial Intelligence Review}}  \bibinfo{volume}{42} (\bibinfo{year}{2014}), \bibinfo{pages}{767--799}.
\newblock


\bibitem[Guo et~al\mbox{.}(2017)]%
        {guo2017deepfm}
\bibfield{author}{\bibinfo{person}{Huifeng Guo}, \bibinfo{person}{Ruiming Tang}, \bibinfo{person}{Yunming Ye}, \bibinfo{person}{Zhenguo Li}, {and} \bibinfo{person}{Xiuqiang He}.} \bibinfo{year}{2017}\natexlab{}.
\newblock \showarticletitle{DeepFM: a factorization-machine based neural network for CTR prediction}.
\newblock \bibinfo{journal}{\emph{arXiv preprint arXiv:1703.04247}} (\bibinfo{year}{2017}).
\newblock


\bibitem[Guo et~al\mbox{.}(2021)]%
        {2021PREFER}
\bibfield{author}{\bibinfo{person}{Y. Guo}, \bibinfo{person}{F. Liu}, \bibinfo{person}{Z. Cai}, \bibinfo{person}{H. Zeng}, {and} \bibinfo{person}{N. Xiao}.} \bibinfo{year}{2021}\natexlab{}.
\newblock \showarticletitle{PREFER: Point-of-interest REcommendation with efficiency and privacy-preservation via Federated Edge leaRning}.
\newblock \bibinfo{journal}{\emph{Proceedings of the ACM on Interactive Mobile Wearable and Ubiquitous Technologies}} \bibinfo{volume}{5}, \bibinfo{number}{1} (\bibinfo{year}{2021}), \bibinfo{pages}{1--25}.
\newblock


\bibitem[Harper and Konstan(2015)]%
        {harper2015movielens}
\bibfield{author}{\bibinfo{person}{F~Maxwell Harper} {and} \bibinfo{person}{Joseph~A Konstan}.} \bibinfo{year}{2015}\natexlab{}.
\newblock \showarticletitle{The movielens datasets: History and context}.
\newblock \bibinfo{journal}{\emph{Acm transactions on interactive intelligent systems (tiis)}} \bibinfo{volume}{5}, \bibinfo{number}{4} (\bibinfo{year}{2015}), \bibinfo{pages}{1--19}.
\newblock


\bibitem[He et~al\mbox{.}(2017)]%
        {he2017neural}
\bibfield{author}{\bibinfo{person}{Xiangnan He}, \bibinfo{person}{Lizi Liao}, \bibinfo{person}{Hanwang Zhang}, \bibinfo{person}{Liqiang Nie}, \bibinfo{person}{Xia Hu}, {and} \bibinfo{person}{Tat-Seng Chua}.} \bibinfo{year}{2017}\natexlab{}.
\newblock \showarticletitle{Neural collaborative filtering}. In \bibinfo{booktitle}{\emph{Proceedings of the 26th international conference on world wide web}}. \bibinfo{pages}{173--182}.
\newblock


\bibitem[Kapoor et~al\mbox{.}(2017)]%
        {kapoor2017review}
\bibfield{author}{\bibinfo{person}{Saakshi Kapoor}, \bibinfo{person}{Vishal Kapoor}, {and} \bibinfo{person}{Rohit Kumar}.} \bibinfo{year}{2017}\natexlab{}.
\newblock \showarticletitle{A REVIEW OF ATTACKS AND ITS DETECTION ATTRIBUTES ON COLLABORATIVE RECOMMENDER SYSTEMS.}
\newblock \bibinfo{journal}{\emph{International Journal of Advanced Research in Computer Science}} \bibinfo{volume}{8}, \bibinfo{number}{7} (\bibinfo{year}{2017}).
\newblock


\bibitem[Kermarrec et~al\mbox{.}(2010)]%
        {kermarrec2010RW}
\bibfield{author}{\bibinfo{person}{Anne-Marie Kermarrec}, \bibinfo{person}{Vincent Leroy}, \bibinfo{person}{Afshin Moin}, {and} \bibinfo{person}{Christopher Thraves}.} \bibinfo{year}{2010}\natexlab{}.
\newblock \showarticletitle{Application of random walks to decentralized recommender systems}. In \bibinfo{booktitle}{\emph{International Conference On Principles Of Distributed Systems}}. Springer, \bibinfo{pages}{48--63}.
\newblock


\bibitem[Long et~al\mbox{.}(2022)]%
        {2022Decentralized}
\bibfield{author}{\bibinfo{person}{J. Long}, \bibinfo{person}{T. Chen}, \bibinfo{person}{Nq~Viet Hung}, {and} \bibinfo{person}{H. Yin}.} \bibinfo{year}{2022}\natexlab{}.
\newblock \showarticletitle{Decentralized Collaborative Learning Framework for Next POI Recommendation}.
\newblock \bibinfo{journal}{\emph{TOIS}} (\bibinfo{year}{2022}).
\newblock


\bibitem[Long et~al\mbox{.}(2023)]%
        {long2023model}
\bibfield{author}{\bibinfo{person}{Jing Long}, \bibinfo{person}{Tong Chen}, \bibinfo{person}{Quoc Viet~Hung Nguyen}, \bibinfo{person}{Guandong Xu}, \bibinfo{person}{Kai Zheng}, {and} \bibinfo{person}{Hongzhi Yin}.} \bibinfo{year}{2023}\natexlab{}.
\newblock \showarticletitle{Model-Agnostic Decentralized Collaborative Learning for On-Device POI Recommendation}. In \bibinfo{booktitle}{\emph{Proceedings of the 46th International ACM SIGIR Conference on Research and Development in Information Retrieval}}. \bibinfo{pages}{423--432}.
\newblock


\bibitem[Long et~al\mbox{.}(2024)]%
        {long2024physical}
\bibfield{author}{\bibinfo{person}{Jing Long}, \bibinfo{person}{Tong Chen}, \bibinfo{person}{Guanhua Ye}, \bibinfo{person}{Kai Zheng}, \bibinfo{person}{Nguyen Quoc~Viet Hung}, {and} \bibinfo{person}{Hongzhi Yin}.} \bibinfo{year}{2024}\natexlab{}.
\newblock \showarticletitle{Physical Trajectory Inference Attack and Defense in Decentralized POI Recommendation}.
\newblock \bibinfo{journal}{\emph{arXiv preprint arXiv:2401.14583}} (\bibinfo{year}{2024}).
\newblock


\bibitem[McAuley et~al\mbox{.}(2015)]%
        {mcauley2015image}
\bibfield{author}{\bibinfo{person}{Julian McAuley}, \bibinfo{person}{Christopher Targett}, \bibinfo{person}{Qinfeng Shi}, {and} \bibinfo{person}{Anton Van Den~Hengel}.} \bibinfo{year}{2015}\natexlab{}.
\newblock \showarticletitle{Image-based recommendations on styles and substitutes}. In \bibinfo{booktitle}{\emph{Proceedings of the 38th international ACM SIGIR conference on research and development in information retrieval}}. \bibinfo{pages}{43--52}.
\newblock


\bibitem[McMahan et~al\mbox{.}(2017)]%
        {pmlr-v54-mcmahan17a}
\bibfield{author}{\bibinfo{person}{Brendan McMahan}, \bibinfo{person}{Eider Moore}, \bibinfo{person}{Daniel Ramage}, \bibinfo{person}{Seth Hampson}, {and} \bibinfo{person}{Blaise Aguera~y Arcas}.} \bibinfo{year}{2017}\natexlab{}.
\newblock \showarticletitle{{Communication-Efficient Learning of Deep Networks from Decentralized Data}}. In \bibinfo{booktitle}{\emph{AISTATS}}. \bibinfo{pages}{1273--1282}.
\newblock


\bibitem[Nasr et~al\mbox{.}(2019)]%
        {nasr2019comprehensive}
\bibfield{author}{\bibinfo{person}{Milad Nasr}, \bibinfo{person}{Reza Shokri}, {and} \bibinfo{person}{Amir Houmansadr}.} \bibinfo{year}{2019}\natexlab{}.
\newblock \showarticletitle{Comprehensive privacy analysis of deep learning: Passive and active white-box inference attacks against centralized and federated learning}. In \bibinfo{booktitle}{\emph{2019 IEEE symposium on security and privacy (SP)}}. IEEE, \bibinfo{pages}{739--753}.
\newblock


\bibitem[Qi et~al\mbox{.}(2017)]%
        {qi2017distributed}
\bibfield{author}{\bibinfo{person}{Lianyong Qi}, \bibinfo{person}{Xuyun Zhang}, \bibinfo{person}{Wanchun Dou}, {and} \bibinfo{person}{Qiang Ni}.} \bibinfo{year}{2017}\natexlab{}.
\newblock \showarticletitle{A distributed locality-sensitive hashing-based approach for cloud service recommendation from multi-source data}.
\newblock \bibinfo{journal}{\emph{IEEE Journal on Selected Areas in Communications}} \bibinfo{volume}{35}, \bibinfo{number}{11} (\bibinfo{year}{2017}), \bibinfo{pages}{2616--2624}.
\newblock


\bibitem[Qu et~al\mbox{.}(2023)]%
        {qu2023semi}
\bibfield{author}{\bibinfo{person}{Liang Qu}, \bibinfo{person}{Ningzhi Tang}, \bibinfo{person}{Ruiqi Zheng}, \bibinfo{person}{Quoc Viet~Hung Nguyen}, \bibinfo{person}{Zi Huang}, \bibinfo{person}{Yuhui Shi}, {and} \bibinfo{person}{Hongzhi Yin}.} \bibinfo{year}{2023}\natexlab{}.
\newblock \showarticletitle{Semi-decentralized federated ego graph learning for recommendation}. In \bibinfo{booktitle}{\emph{Proceedings of the ACM Web Conference 2023}}. \bibinfo{pages}{339--348}.
\newblock


\bibitem[Qu et~al\mbox{.}(2024b)]%
        {qu2024towards}
\bibfield{author}{\bibinfo{person}{Liang Qu}, \bibinfo{person}{Wei Yuan}, \bibinfo{person}{Ruiqi Zheng}, \bibinfo{person}{Lizhen Cui}, \bibinfo{person}{Yuhui Shi}, {and} \bibinfo{person}{Hongzhi Yin}.} \bibinfo{year}{2024}\natexlab{b}.
\newblock \showarticletitle{Towards Personalized Privacy: User-Governed Data Contribution for Federated Recommendation}.
\newblock \bibinfo{journal}{\emph{arXiv preprint arXiv:2401.17630}} (\bibinfo{year}{2024}).
\newblock


\bibitem[Qu et~al\mbox{.}(2024a)]%
        {qu2024budgeted}
\bibfield{author}{\bibinfo{person}{Yunke Qu}, \bibinfo{person}{Tong Chen}, \bibinfo{person}{Quoc Viet~Hung Nguyen}, {and} \bibinfo{person}{Hongzhi Yin}.} \bibinfo{year}{2024}\natexlab{a}.
\newblock \showarticletitle{Budgeted embedding table for recommender systems}. In \bibinfo{booktitle}{\emph{Proceedings of the 17th ACM International Conference on Web Search and Data Mining}}. \bibinfo{pages}{557--566}.
\newblock


\bibitem[Rendle et~al\mbox{.}(2012)]%
        {rendle2012bpr}
\bibfield{author}{\bibinfo{person}{Steffen Rendle}, \bibinfo{person}{Christoph Freudenthaler}, \bibinfo{person}{Zeno Gantner}, {and} \bibinfo{person}{Lars Schmidt-Thieme}.} \bibinfo{year}{2012}\natexlab{}.
\newblock \showarticletitle{BPR: Bayesian personalized ranking from implicit feedback}.
\newblock \bibinfo{journal}{\emph{arXiv preprint arXiv:1205.2618}} (\bibinfo{year}{2012}).
\newblock


\bibitem[Rong et~al\mbox{.}(2022a)]%
        {rong2022poisoning}
\bibfield{author}{\bibinfo{person}{Dazhong Rong}, \bibinfo{person}{Qinming He}, {and} \bibinfo{person}{Jianhai Chen}.} \bibinfo{year}{2022}\natexlab{a}.
\newblock \showarticletitle{Poisoning deep learning based recommender model in federated learning scenarios}.
\newblock \bibinfo{journal}{\emph{arXiv preprint arXiv:2204.13594}} (\bibinfo{year}{2022}).
\newblock


\bibitem[Rong et~al\mbox{.}(2022b)]%
        {rong2022fedrecattack}
\bibfield{author}{\bibinfo{person}{Dazhong Rong}, \bibinfo{person}{Shuai Ye}, \bibinfo{person}{Ruoyan Zhao}, \bibinfo{person}{Hon~Ning Yuen}, \bibinfo{person}{Jianhai Chen}, {and} \bibinfo{person}{Qinming He}.} \bibinfo{year}{2022}\natexlab{b}.
\newblock \showarticletitle{FedRecAttack: model poisoning attack to federated recommendation}. In \bibinfo{booktitle}{\emph{2022 IEEE 38th International Conference on Data Engineering (ICDE)}}. IEEE, \bibinfo{pages}{2643--2655}.
\newblock


\bibitem[Sun et~al\mbox{.}(2022)]%
        {sun2022survey}
\bibfield{author}{\bibinfo{person}{Zehua Sun}, \bibinfo{person}{Yonghui Xu}, \bibinfo{person}{Yong Liu}, \bibinfo{person}{Wei He}, \bibinfo{person}{Yali Jiang}, \bibinfo{person}{Fangzhao Wu}, {and} \bibinfo{person}{Lizhen Cui}.} \bibinfo{year}{2022}\natexlab{}.
\newblock \showarticletitle{A Survey on Federated Recommendation Systems}.
\newblock \bibinfo{journal}{\emph{arXiv preprint arXiv:2301.00767}} (\bibinfo{year}{2022}).
\newblock


\bibitem[Tolpegin et~al\mbox{.}(2020)]%
        {tolpegin2020data}
\bibfield{author}{\bibinfo{person}{Vale Tolpegin}, \bibinfo{person}{Stacey Truex}, \bibinfo{person}{Mehmet~Emre Gursoy}, {and} \bibinfo{person}{Ling Liu}.} \bibinfo{year}{2020}\natexlab{}.
\newblock \showarticletitle{Data poisoning attacks against federated learning systems}. In \bibinfo{booktitle}{\emph{Computer Security--ESORICS 2020: 25th European Symposium on Research in Computer Security, ESORICS 2020, Guildford, UK, September 14--18, 2020, Proceedings, Part I 25}}. Springer, \bibinfo{pages}{480--501}.
\newblock


\bibitem[Wang et~al\mbox{.}(2017)]%
        {wang2017location}
\bibfield{author}{\bibinfo{person}{Hao Wang}, \bibinfo{person}{Yanmei Fu}, \bibinfo{person}{Qinyong Wang}, \bibinfo{person}{Hongzhi Yin}, \bibinfo{person}{Changying Du}, {and} \bibinfo{person}{Hui Xiong}.} \bibinfo{year}{2017}\natexlab{}.
\newblock \showarticletitle{A location-sentiment-aware recommender system for both home-town and out-of-town users}. In \bibinfo{booktitle}{\emph{Proceedings of the 23rd ACM SIGKDD international conference on knowledge discovery and data mining}}. \bibinfo{pages}{1135--1143}.
\newblock


\bibitem[Wang et~al\mbox{.}(2020)]%
        {2020Next}
\bibfield{author}{\bibinfo{person}{Q. Wang}, \bibinfo{person}{H. Yin}, \bibinfo{person}{T. Chen}, \bibinfo{person}{Z. Huang}, {and} \bibinfo{person}{Nqv Hung}.} \bibinfo{year}{2020}\natexlab{}.
\newblock \showarticletitle{Next Point-of-Interest Recommendation on Resource-Constrained Mobile Devices}. In \bibinfo{booktitle}{\emph{WWW '20: The Web Conference 2020}}.
\newblock


\bibitem[Wu et~al\mbox{.}(2022)]%
        {wu2022fedattack}
\bibfield{author}{\bibinfo{person}{Chuhan Wu}, \bibinfo{person}{Fangzhao Wu}, \bibinfo{person}{Tao Qi}, \bibinfo{person}{Yongfeng Huang}, {and} \bibinfo{person}{Xing Xie}.} \bibinfo{year}{2022}\natexlab{}.
\newblock \showarticletitle{FedAttack: Effective and covert poisoning attack on federated recommendation via hard sampling}. In \bibinfo{booktitle}{\emph{SIGKDD}}. \bibinfo{pages}{4164--4172}.
\newblock


\bibitem[Xia et~al\mbox{.}(2023)]%
        {xia2023efficient}
\bibfield{author}{\bibinfo{person}{Xin Xia}, \bibinfo{person}{Junliang Yu}, \bibinfo{person}{Qinyong Wang}, \bibinfo{person}{Chaoqun Yang}, \bibinfo{person}{Nguyen Quoc~Viet Hung}, {and} \bibinfo{person}{Hongzhi Yin}.} \bibinfo{year}{2023}\natexlab{}.
\newblock \showarticletitle{Efficient on-device session-based recommendation}.
\newblock \bibinfo{journal}{\emph{ACM Transactions on Information Systems}} \bibinfo{volume}{41}, \bibinfo{number}{4} (\bibinfo{year}{2023}), \bibinfo{pages}{1--24}.
\newblock


\bibitem[Yang et~al\mbox{.}(2020)]%
        {yang2020federated}
\bibfield{author}{\bibinfo{person}{Liu Yang}, \bibinfo{person}{Ben Tan}, \bibinfo{person}{Vincent~W Zheng}, \bibinfo{person}{Kai Chen}, {and} \bibinfo{person}{Qiang Yang}.} \bibinfo{year}{2020}\natexlab{}.
\newblock \showarticletitle{Federated recommendation systems}.
\newblock In \bibinfo{booktitle}{\emph{Federated Learning}}. \bibinfo{publisher}{Springer}, \bibinfo{pages}{225--239}.
\newblock


\bibitem[Yang et~al\mbox{.}(2022)]%
        {yang2022dpmf}
\bibfield{author}{\bibinfo{person}{Xu Yang}, \bibinfo{person}{Yuchuan Luo}, \bibinfo{person}{Shaojing Fu}, \bibinfo{person}{Ming Xu}, {and} \bibinfo{person}{Yingwen Chen}.} \bibinfo{year}{2022}\natexlab{}.
\newblock \showarticletitle{DPMF: Decentralized Probabilistic Matrix Factorization for Privacy-Preserving Recommendation}.
\newblock \bibinfo{journal}{\emph{Applied Sciences}} \bibinfo{volume}{12}, \bibinfo{number}{21} (\bibinfo{year}{2022}), \bibinfo{pages}{11118}.
\newblock


\bibitem[Yao et~al\mbox{.}(2021)]%
        {yao2021device}
\bibfield{author}{\bibinfo{person}{Jiangchao Yao}, \bibinfo{person}{Feng Wang}, \bibinfo{person}{KunYang Jia}, \bibinfo{person}{Bo Han}, \bibinfo{person}{Jingren Zhou}, {and} \bibinfo{person}{Hongxia Yang}.} \bibinfo{year}{2021}\natexlab{}.
\newblock \showarticletitle{Device-Cloud Collaborative Learning for Recommendation}.
\newblock \bibinfo{journal}{\emph{arXiv preprint arXiv:2104.06624}} (\bibinfo{year}{2021}).
\newblock


\bibitem[Ye et~al\mbox{.}(2022)]%
        {2022YE}
\bibfield{author}{\bibinfo{person}{Guanhua Ye}, \bibinfo{person}{Hongzhi Yin}, {and} \bibinfo{person}{Tong Chen}.} \bibinfo{year}{2022}\natexlab{}.
\newblock \showarticletitle{A Decentralized Collaborative Learning Framework Across Heterogeneous Devices for Personalized Predictive Analytics}.
\newblock  (\bibinfo{year}{2022}).
\newblock


\bibitem[Yin et~al\mbox{.}(2018)]%
        {yin2018byzantine}
\bibfield{author}{\bibinfo{person}{Dong Yin}, \bibinfo{person}{Yudong Chen}, \bibinfo{person}{Ramchandran Kannan}, {and} \bibinfo{person}{Peter Bartlett}.} \bibinfo{year}{2018}\natexlab{}.
\newblock \showarticletitle{Byzantine-robust distributed learning: Towards optimal statistical rates}. In \bibinfo{booktitle}{\emph{International Conference on Machine Learning}}. PMLR, \bibinfo{pages}{5650--5659}.
\newblock


\bibitem[Yin et~al\mbox{.}(2024)]%
        {yin2024ondevice}
\bibfield{author}{\bibinfo{person}{Hongzhi Yin}, \bibinfo{person}{Liang Qu}, \bibinfo{person}{Tong Chen}, \bibinfo{person}{Wei Yuan}, \bibinfo{person}{Ruiqi Zheng}, \bibinfo{person}{Jing Long}, \bibinfo{person}{Xin Xia}, \bibinfo{person}{Yuhui Shi}, {and} \bibinfo{person}{Chengqi Zhang}.} \bibinfo{year}{2024}\natexlab{}.
\newblock \bibinfo{title}{On-Device Recommender Systems: A Comprehensive Survey}.
\newblock
\newblock
\showeprint[arxiv]{2401.11441}~[cs.IR]


\bibitem[Yuan et~al\mbox{.}(2023a)]%
        {yuan2023interaction}
\bibfield{author}{\bibinfo{person}{Wei Yuan}, \bibinfo{person}{Chaoqun Yang}, \bibinfo{person}{Quoc Viet~Hung Nguyen}, \bibinfo{person}{Lizhen Cui}, \bibinfo{person}{Tieke He}, {and} \bibinfo{person}{Hongzhi Yin}.} \bibinfo{year}{2023}\natexlab{a}.
\newblock \showarticletitle{Interaction-level membership inference attack against federated recommender systems}.
\newblock \bibinfo{journal}{\emph{arXiv preprint arXiv:2301.10964}} (\bibinfo{year}{2023}).
\newblock


\bibitem[Yuan et~al\mbox{.}(2023b)]%
        {yuan2023federated}
\bibfield{author}{\bibinfo{person}{Wei Yuan}, \bibinfo{person}{Hongzhi Yin}, \bibinfo{person}{Fangzhao Wu}, \bibinfo{person}{Shijie Zhang}, \bibinfo{person}{Tieke He}, {and} \bibinfo{person}{Hao Wang}.} \bibinfo{year}{2023}\natexlab{b}.
\newblock \showarticletitle{Federated unlearning for on-device recommendation}. In \bibinfo{booktitle}{\emph{Proceedings of the Sixteenth ACM International Conference on Web Search and Data Mining}}. \bibinfo{pages}{393--401}.
\newblock


\bibitem[Yuan et~al\mbox{.}(2023c)]%
        {yuan2023manipulating}
\bibfield{author}{\bibinfo{person}{Wei Yuan}, \bibinfo{person}{Shilong Yuan}, \bibinfo{person}{Chaoqun Yang}, \bibinfo{person}{Nguyen Quoc Viet~hung}, {and} \bibinfo{person}{Hongzhi Yin}.} \bibinfo{year}{2023}\natexlab{c}.
\newblock \showarticletitle{Manipulating Visually Aware Federated Recommender Systems and Its Countermeasures}.
\newblock \bibinfo{journal}{\emph{ACM Transactions on Information Systems}} \bibinfo{volume}{42}, \bibinfo{number}{3} (\bibinfo{year}{2023}), \bibinfo{pages}{1--26}.
\newblock


\bibitem[Zhang et~al\mbox{.}(2021b)]%
        {zhang2021survey}
\bibfield{author}{\bibinfo{person}{Chen Zhang}, \bibinfo{person}{Yu Xie}, \bibinfo{person}{Hang Bai}, \bibinfo{person}{Bin Yu}, \bibinfo{person}{Weihong Li}, {and} \bibinfo{person}{Yuan Gao}.} \bibinfo{year}{2021}\natexlab{b}.
\newblock \showarticletitle{A survey on federated learning}.
\newblock \bibinfo{journal}{\emph{Knowledge-Based Systems}}  \bibinfo{volume}{216} (\bibinfo{year}{2021}), \bibinfo{pages}{106775}.
\newblock


\bibitem[Zhang et~al\mbox{.}(2021a)]%
        {zhang2021double}
\bibfield{author}{\bibinfo{person}{Junwei Zhang}, \bibinfo{person}{Min Gao}, \bibinfo{person}{Junliang Yu}, \bibinfo{person}{Lei Guo}, \bibinfo{person}{Jundong Li}, {and} \bibinfo{person}{Hongzhi Yin}.} \bibinfo{year}{2021}\natexlab{a}.
\newblock \showarticletitle{Double-scale self-supervised hypergraph learning for group recommendation}. In \bibinfo{booktitle}{\emph{Proceedings of the 30th ACM international conference on information \& knowledge management}}. \bibinfo{pages}{2557--2567}.
\newblock


\bibitem[Zhang et~al\mbox{.}(2021c)]%
        {zhang2021pipattack}
\bibfield{author}{\bibinfo{person}{Shijie Zhang}, \bibinfo{person}{Hongzhi Yin}, \bibinfo{person}{Tong Chen}, \bibinfo{person}{Zi Huang}, \bibinfo{person}{Quoc Viet~Hung Nguyen}, {and} \bibinfo{person}{Lizhen Cui}.} \bibinfo{year}{2021}\natexlab{c}.
\newblock \showarticletitle{PipAttack: Poisoning Federated Recommender Systems forManipulating Item Promotion}.
\newblock \bibinfo{journal}{\emph{arXiv preprint arXiv:2110.10926}} (\bibinfo{year}{2021}).
\newblock


\bibitem[Zhang et~al\mbox{.}(2022)]%
        {zhang2022pipattack}
\bibfield{author}{\bibinfo{person}{Shijie Zhang}, \bibinfo{person}{Hongzhi Yin}, \bibinfo{person}{Tong Chen}, \bibinfo{person}{Zi Huang}, \bibinfo{person}{Quoc Viet~Hung Nguyen}, {and} \bibinfo{person}{Lizhen Cui}.} \bibinfo{year}{2022}\natexlab{}.
\newblock \showarticletitle{Pipattack: Poisoning federated recommender systems for manipulating item promotion}. In \bibinfo{booktitle}{\emph{Proceedings of the Fifteenth ACM International Conference on Web Search and Data Mining}}. \bibinfo{pages}{1415--1423}.
\newblock


\bibitem[Zhao et~al\mbox{.}(2023)]%
        {zhao2023survey}
\bibfield{author}{\bibinfo{person}{Wayne~Xin Zhao}, \bibinfo{person}{Kun Zhou}, \bibinfo{person}{Junyi Li}, \bibinfo{person}{Tianyi Tang}, \bibinfo{person}{Xiaolei Wang}, \bibinfo{person}{Yupeng Hou}, \bibinfo{person}{Yingqian Min}, \bibinfo{person}{Beichen Zhang}, \bibinfo{person}{Junjie Zhang}, \bibinfo{person}{Zican Dong}, {et~al\mbox{.}}} \bibinfo{year}{2023}\natexlab{}.
\newblock \showarticletitle{A survey of large language models}.
\newblock \bibinfo{journal}{\emph{arXiv preprint arXiv:2303.18223}} (\bibinfo{year}{2023}).
\newblock


\bibitem[Zheng et~al\mbox{.}(2024)]%
        {zheng2024decentralized}
\bibfield{author}{\bibinfo{person}{Ruiqi Zheng}, \bibinfo{person}{Liang Qu}, \bibinfo{person}{Tong Chen}, \bibinfo{person}{Lizhen Cui}, \bibinfo{person}{Yuhui Shi}, {and} \bibinfo{person}{Hongzhi Yin}.} \bibinfo{year}{2024}\natexlab{}.
\newblock \showarticletitle{Decentralized Collaborative Learning with Adaptive Reference Data for On-Device POI Recommendation}.
\newblock \bibinfo{journal}{\emph{arXiv preprint arXiv:2401.13448}} (\bibinfo{year}{2024}).
\newblock


\end{thebibliography}

\appendix

\end{document}